\documentstyle[a4]{article}
\begin{document}

\newcommand{\be}{\begin{equation}}
\newcommand{\ee}{\end{equation}}
\newcommand{\epe}{\end{equation}}
\newcommand{\bea}{\begin{eqnarray}}
\newcommand{\eea}{\end{eqnarray}}
\newcommand{\ba}{\begin{eqnarray*}}
\newcommand{\ea}{\end{eqnarray*}}
\newcommand{\epa}{\end{eqnarray*}}
\newcommand{\ar}{\rightarrow}

\def\s{\sigma}
\def\r{\rho}
\def\D{\Delta}
\def\R{I\!\!R}
\def\l{\lambda}
\def\g{\gamma}
\def\D{\Delta}
\def\cD{{\cal D}}
\def\cH{{\cal H}}
\def\dA{A^{\dag}}
\def\d{\delta}
\def\T{\tilde{T}}
\def\k{\kappa}
\def\t{\tau}
\def\f{\phi}
\def\p{\psi}
\def\pmh{{in/out}}
\def\inn{{in}}
\def\out{{out}}
\def\z{\zeta}
\def\ep{\epsilon}
\def\hx{\widehat{\xi}}
\def\a{\alpha}
\def\b{\beta}
\def\O{\Omega}
\def\H{\cal H}
\def\M{\cal M}
\def\g{\hat g}
\newcommand{\dslash}{\partial\!\!\!/}
\newcommand{\aslash}{a\!\!\!/}
\newcommand{\eslash}{e\!\!\!/}
\newcommand{\bslash}{b\!\!\!/}
\newcommand{\vslash}{v\!\!\!/}
\newcommand{\rslash}{r\!\!\!/}
\newcommand{\cslash}{c\!\!\!/}
\newcommand{\fslash}{f\!\!\!/}
\newcommand{\Dslash}{D\!\!\!\!/}
\newcommand{\Aslash}{{\cal A}\!\!\!\!/}


\begin{center}

{\large Strings in Horizons, Dissipation and a Possible
Interpretation of the Hagedorn Temperature.}

\vspace*{1.2cm} {\large M. C. B. Abdalla $^{\dag,}$\footnote{e-mail:
mabdalla@ift.unesp.br}, M. Botta Cantcheff $^{\ddag}$
\footnote{e-mail: botta@cbpf.br}, Daniel L. Nedel $^{\dag}$
$^{\star,}$\footnote{e-mail: dnedel.unipampa@ufpel.edu.br}}

\vspace*{3mm} $^{\dag}$ Instituto de F\'{\i}sica Te\'orica
(IFT/UNESP)

Rua Pamplona, 145 - Bela Vista

01405-900 - S\~ao Paulo, SP - Brasil.

\vspace{3mm}

$^{\ddag}$ Instituto de Fisica La Plata, CONICET, UNLP \\
CC 67, Calles 49 y 115, 1900 La Plata, Buenos Aires, Argentina

\vspace{3mm}

$^{\star}$UFPEL/UNIPAMPA-Bag\'e \\
Rua Carlos Barbosa S/Nº,
Bairro Get\'ulio Vargas \\
 96412-420 Bag\'e, RS, Brasil

\end{center}

\begin{abstract}
\noindent

We consider the entanglement of closed bosonic strings intersecting
the event horizon of a Rindler spacetime and, by using some
simplified (rather semiclassical) arguments and some elements of the
string field theory, we show the existence of a critical temperature
beyond which closed strings \emph{cannot be in thermal equilibrium}.
The order of magnitude of this critical value coincides with the
Hagedorn temperature, which suggests an interpretation consistent
with the fact of having a partition function which is bad defined
for temperatures higher than it. Possible implications of the
present approach on the microscopical structure of stretched
horizons are also pointed out.

\end{abstract}

\section{Introduction}

 A central issue in string theory at finite
 temperature is the meaning of the Hagedorn temperature \cite{witt}, which may be related to a critical
  acceleration via Unruh effect. Some authors argued that strings sufficiently near of
   an event horizon should be accelerated so much that this critical value would be exceeded,
    and an infinite energy would be required for this purpose; consequently, it is interpreted that
     the string would slip into the Black Hole \cite{slips}. The main
     question we are addressing  is about
      the right way of describing this process.

Furthermore, on the other hand, a Black Hole interacts with systems which are at
rest with respect to
 them (called fiducial systems) as a {\it dissipative} effective membrane placed on the horizon neighborhood.
  This idea was first used in astrophysics for a long time and referred to
  as the ``membrane paradigm" \cite{price}.
   More recently, it was revisited in the string theory context with the suggestive name of
    ``stretched horizons"  \cite{suss1} with an important interest
     in its actual physical meaning and microscopical picture.
We are going to see here that these two issues result to be related
      in the context of strings.

 In fact, stretched horizons should be placed at a distance from the event
horizon, related to the $l_s$ string length scale \cite{suss2}.
According to some references (e. g. \cite{slips}, \cite{sanchez2}),
the string is maximally accelerated at this distance, and the
corresponding Unruh temperature is associated to the Hagedorn
temperature. Here we propose that in this situation the event
horizon intersects the worldsheet in order to show that, at the
 corresponding critical temperature/acceleration value, the string vacuum (seen by
accelerated observers) is an entangled state of string modes living in
the two causally disconnected regions.
Notice that this \textit{configuration} may be seen
as two open strings ending precisely at the horizon and interpreted
as a vertex diagram. Finally, we are going to argue that when
a second quantized string theory is taken into account, the
corresponding  Hamiltonian to such a configuration (vertex) of the
closed string intersected by the event horizon is actually related
to dissipative processes when a string field description is adopted.
Some results recently reported show that the infrared behavior of
theories whose dual bulk-gravities contain a black brane is governed
by hydrodynamics \cite{bb1,bb2,bb3}. The most interesting result in
this sense is the discovery of a universal value for the ratio of
shear viscosity to entropy density \cite{universal}. So, although we
are considering a Rindler spacetime, the fact that the
configurations of string at the horizon produce a ``universal"
dissipative Hamiltonian for open string might reveal something about
the microscopic nature of this universal hydrodynamic behavior from
the string's perspective.

The paper is organized as follows. In Section 2 we study the
entanglement of the string vacuum and point out its relation with
the microscopical structure of the stretched horizon. We use some
semiclassical arguments to show the existence of
 a critical temperature  where these states arise. In
Section 3 we use some elementary issues of light-cone string field
theory to argue for the existence of dissipative process beyond that
critical temperature. In Section 4,  we summarize some concluding
remarks where briefly discuss  the possibility of identifying that
critical value with the Hagedorn temperature (whose order of
magnitude are in agreement), and its eventual significance in order
to explain the singular behavior of the partition function beyond
this value.

\section{Entanglement of closed string in Rindler horizons and critical acceleration/temperature.}

Let us consider a bosonic closed string in light-cone gauge (LCG).
The general solution  for the transversal (physical)
 degrees of freedom is
\be\label{stringsol} X^i(t,\sigma)= x^i + 2\a' p^i t + i
\sqrt{\a'/2} \sum_{n \neq 0} \left(\a_n^i e^{-2in(t-\s)} +
\b_n^i e^{-2in(t + \s)}\right)/n \ee where $\s\in [0 , 2\pi] \sim
S^1$ , $t\in \R$, the worldsheet manifold is $W\sim S^1
\times \R_t$ and the worldsheet metric components corresponding to these coordinates are given by $g_0 = diag (-1,1)$. The field $X$ defines an embedding of $W$ into the
spacetime $M$.
 In the  presence of event horizons,
  the spacetime may be divided in two (for simplicity)
   regions denoted by $M^\pmh$, causally disconnected/connected with certain class of observers.

\vspace{0.2cm}

The key idea in this simplified model is that, at least in a
classical (and even semiclassical) sense, when the closed string
(thought here as a classical extended object) intersects an event
horizon at some point, no microscopical information goes through it, 
and this constitutes a critical regime. In this situation string
vibrations are more and more red shifted so as they approximate to
the point of intersection with the horizon, and are frozen at this
point \cite{sanchez-vega}. Let us consider now a convenient mathematical approach to this scenario.
 \vspace{0.4cm}

  For accelerated observers in the spacetime,
the spacetime metric reads \be\label{rindlertarget} dS^2= -R^2 dT^2
+ dR^2 + dX_i  dX^i \ee which is called the Rindler metric. On the
other hand, 
the spatial coordinate $R$ may be expressed as \be R^2 =
X^+ X^- \ee in terms of the light cone coordinates.

The Rindler spacetime is simply the wedge $X^+ > 0 , X^- > 0$ of the Minkowski spacetime, and its boundary (horizon) is given by the surfaces $X^+ = 0 $ and $ X^- = 0$ \footnote{We refer to this as $M^{out}$}.
A Rindler observer, following a timelike curve in one of these four regions (say, the wedge I) in which the Minkowski spacetime is divided, only detects microscopic information coming from this one. In fact, the horizon surfaces correspond to infinite values of the proper time of these uniformly accelerated  observers, $T(X^\pm = 0)= \pm \infty$ . So despite information may fall into the wedge II (in the causal future of the wedge I) crossing the null line $X^+ = 0 $, it is not effectively observed by these Rindler observers in a finite proper time. In this precise sense one may refer to causal independence between the wedges.

The light
cone gauge fixing consists in taking $t=X^\pm$ as time parameter of
the string (LCG) . In particular, if $t=X^+$ (notice that $t=X^-$ is an
equivalent choice) we get \be R^2 = t X^- \ee which shows that the
(closed) string intersects the horizon at least in $t=0$ (so as in
worldsheet points such that $X^- =0$). So, if the string worldsheet
is $W\sim S^1 \times \R_t$, there is a complete circle where it
intersects the horizon. This shows our initial assertion on the
splitting in two regions $W^\pmh$. Thus, in agreement with Refs.
\cite{Israel} \cite{laflamme}, the string lying behind the horizon
$t<0$ affects the string state by entanglement. So, by requiring
continuity of the embedding $X: W \to M $, the string states in the
horizon are given by the gluing condition \be (X_\inn^i - X_\out^i
)|_{t=0} |h \rangle = 0 \label{gluing}\ee which actually constitutes
a boundary state since it is implemented at $t=0$. By using the
expansion in modes (\ref{stringsol}) we may write \be\label{B1}
\b^{\inn \,,\,i}_n - e^{i\,2n\,\lambda}\;\a^{\out \,,\,i}_n \;|h
\rangle = 0, \ee

\be \label{B2} \a^{\inn \,,\,i}_n - e^{-i\,2n\,\lambda}\;\b^{\out
\,,\,i}_n \;|h \rangle = 0, \ee

 \be \label{B3}
  x^{\inn \,,\, i}_0 - x^{\out \,, \,i}_0 |h \rangle = 0 ,\ee where $x^{\inn \, ,\,i}_0,
x^{\out \,,\, i}_0$ denote both center of mass coordinates. We have
defined the respective string coordinates up to a shift: $ \s_\out
\equiv \s_\inn + \lambda$, so $\lambda$ is the relative twist
between the respective ends of the gluing strings
\cite{gluing-thorus} \footnote{The signal of $t$ is inverted in the
horizon so, in order to define both fields $X^\pmh$ as
parameterized by a positive time parameter, it is required that
right($\pmh$)/left($\pmh$) modes be defined such that they
contribute both to the same term in the expansion of the general
solution of eq. (\ref{gluing}).}. The solution of
(\ref{B1},\ref{B2},\ref{B3}) may be expressed then as:
\be\label{boundary} |h \rangle  = N_H \; \delta(x^{\inn \, ,\,i}_0 -
x^{\out \,, \,i}_0 ) \prod_{n>0 , i} \,\, e^{q_n \a^{\inn \,,\,
i}_{-n} \b^{\out \,,\, i}_{-n} }\,\,
 \prod_{n>0 , i} \,\,
e^{{\bar q_n } \b^{\inn \,, \,i}_{-n} \a^{\out\, ,\, i}_{-n} } |0
\rangle \ee
where $q_n = e^{i \,2n \,\lambda}$ and $N_H$ is the
proper normalizing constant.

Notice that in addition to the circle $t=0$ , the string intersects
the horizon at the points $X^- =0 $. Classically, this condition
defines a curve on the worldsheet $\s_H = S(t)$ for $t>0$ which
clearly separates the worldsheet in two parts, with topologies $[0,
S(t)]\times \R_{t>0}$ and $[S(t), 2\pi]\times \R_{t>0}$
respectively, which are open sheets. So, in other words, if the
world tube goes through the axis $X^- =0 $ along the time $X^+$, the
worldsheet points are separated in two sets $W^\pmh$ defined
such that the map  $X^-(p \in W^\pmh)$ is positive/negative, respectively.
In order to obtain the quantum states corresponding to this
condition, once again, one might take the other possible choice
consistent with the light cone gauge (LCG), namely $X^- =t$, which
clearly provides the same state (\ref{boundary}) \footnote{ Notice
that by quantizing the condition $X^- =0 $, thought $X^- $ as the
time parameter, corresponds to view the interval $[0, S(t)]$ as the
evolution time rather than the (open) string range of the coordinate
$\s$. This resembles the transformation which interchanges closed
into open channel.}. The study of the quantum solutions of the
condition $X^- |h \rangle=0$ is an interesting issue in itself and
will be investigated in another paper. However here, we are focused
in showing that this situation may be described as a genuine
entanglement between string degrees of freedom, viewed as a two
dimensional field theory.

Let us study then the closed string entangled that we are talking
about. From a strictly topological point of view, a closed string
may be separated by the event horizon in two physical regions (simply connected or not),
$W^{in/out} \; / \; X(W^\pmh) = M^\pmh$, which
 are clearly {\it open}.
 
 Because this is a local free
    theory, ${\H}_{closed}$ can be trivially separated in the direct product of two independent Hilbert
     spaces ${\cal H} [ S^{in} ] \otimes {\cal H}[S^{out}]$ where $ S^{in} \cup S^{out} = S^1$.
To avoid an (in principle) arbitrary confusion with standard open
strings (defined by Dirichlet/Neumann combinations of boundary
conditions) we name them ``c-open strings". In fact we can define
the fields $X^\pmh$ as the restriction of the field $X$ to each
wedge $W^\pmh$ and quantize them with boundary matching conditions,
$X^\out |_{\partial S^\out }= X^{in } |_{\partial S^\inn }$. The
solutions simply are the restriction
 of the solution (\ref{stringsol}) to the respective intervals. So, the vacuum state of a closed
  string may be written as a tensor product of two states in ${\cal H}^{in} \otimes {\cal H} ^{out}$, namely $|0_\inn \rangle|0_\out \rangle$.
   Then, it is evident that in this situation the local degrees of freedom of the part of the string behind the horizon
    ($W^\inn$) produce {\it entanglement} and the fundamental state of the system is a mixed/entangled
     state which clearly differs from the no-entangled vacuum state where all the local degrees of
      freedom of the closed string are causally connected
 \footnote{We remarkably notice that in principle, an arbitrary separation of a closed string in two c-open strings
may always be considered, but it is merely a formal construction
except in the situation where a closed string intersects a horizon.
This will become clearer in the last Section when we describe
this as a vertex whose contribution shall be irrelevant except in
presence of the horizon.}.

 \vspace{0.4cm}
 
So in this sense, we wish to show that 
the problem may be handled as in the Unruh-Hawking effect, where an ordinary field theory is quantized on a two-dimensional base manifold (the world-sheet) with an event horizon (see Ref.\cite{vitiello}). We are going to show indeed that when a closed string intersects the event horizon of accelerated observers, the string vacuum  coincides with the boundary state (\ref{boundary}) which entangles inner and outer string modes, and it is related to the inertial vacuum state through a Bogoliubov transformation.

To this purpose we may observe that the embedding $X:W\to M$, assumed to be smooth, induces a compatible
worldsheet metric through the expression
\be
 g_{(W) ab} = \partial_a X^\mu\, \partial_b X^\nu \, g_{(M) \mu \nu}.
 \ee
 In particular, it is straightforward to verify that the worldsheet element of distance induced from (\ref{rindlertarget}) reads\footnote{For more details, see Appendix.} 
 \be\label{rindler} ds^2 = - r^2 d\tau^2 + dr^2 ,\ee
 where the vector field $\tau^a := \nabla^a \tau \,\, \,(a = 0, 1)$ in the worldsheet is related to 
 the $d$-vector tangent to the congruence of 
accelerated curves (Rindler observers), 
$\tau^\mu  \equiv \nabla^\mu T, \,\, \,(\mu = 0..d)$, 
through the expression.

\be \label{emb}\tau^\mu = \partial_a
X^\mu \tau^a ,\ee given by the pull-back corresponding to the string embedding. 
Similarly, we have defined the parameter $r^a :=\nabla^a r$ as related to the Rindler's spatial coordinate $R_\mu :=\partial_\mu R$
through the pull-back
\be \label{embR}R^\mu = r^a \partial_a X^\mu  .\ee 
This actually describes the
worldsheet geometry in a neighborhood of the horizon.

Notice that the worldsheet horizon, defined by $r=0$, actually
describes the intersection of the string with a
   horizon in {\it the space time} for a time string parameter
    $\tau$  chosen in agreement with the proper time of certain congruence of
     time like curves in the spacetime associated with a
 particular class of (properly accelerated) observers.
If $p\in H_W$, the time-time component of $g_{(W)}$ vanishes at this
 point, then (since
   the pull-back is well defined the character -time like- of a vector is
   preserved by this) automatically  the time-time component of $g_{(M)}$ also vanishes at $X(p)$,
\be
 0 = g_{(W) \tau \tau}|_{p } = g_{(W) ab} \,\tau^a \tau^b |_{p }=   
  g_{(M) \mu \nu}|_{X(p) }\,(\partial_a X^\mu \tau^a |_{p })\, (\partial_b X^\nu\,\tau^b |_{p }) =  g_{(M) \tau \tau}|_{X(p)} \,\, ,  
 \ee
 then $X(p)\in H_M$, the horizon of the observers that follow time-like curves generated by $\tau^\mu$. The same statement holds in the inverse sense, and therefore we have that: The horizons of the respective congruences,  related by (\ref{emb}), are such that
 $p\in H_W$ if and only if $X(p)\in H_M$. However, let us remark that the presence of $H_W$ does not imply a coincident $H_M$, unless do exist spacetime observers whose proper time are \emph{synchronized} with the time string parameter (as it is the case of the particular class of Rindler observers we are considering in this framework) \footnote{In Ref. \cite{rev} there are examples where worldsheet horizons may not coincide
with spacetime horizon crossings.}. This is a useful property for the situation we
     are analyzing and simplifies considerably the analysis, since it allows us to deal with the problem of describing
      the string intersecting an event horizon directly in the {\it worldsheet},
       rather than imposing this condition
in the target space \footnote{ In other words, the string
coordinates are {\it operators} which should be equaled to the
horizon coordinates, which are c-numbers. So this should be
implemented on states rather than on operators, which should be
similar to (\ref{gluing})}. Thus, it is actually natural to interpret
the relation (\ref{emb}) in a semiclassical sense (taking 
expectation value in the right hand side), since the target vector
$\tau^\mu $ refers to the $d+1$-velocity of {\it classical}
observers. One may therefore define classical frames (associated to observers in the spacetime) 
such they detect the horizon $H_M \, \, \,/ \,H_M \cap X(W) = X(H_W)$ in this way.

 This picture is similar to the Unruh-Hawking scenario and we may follow the standard procedure, consisting in
 quantizing $D-2$ ordinary scalar fields on a two
dimensional (lorentzian) manifold $W$ in both accelerated and
inertial frames and find out the Bogoliubov transformation which relates the respective Fock spaces.

\vspace{0.6cm}

In the LCG, the physical degrees of freedom dynamics for closed
bosonic string is governed by the equation \be\label{KG-LCG}\Box
\,\, X^i \,= 0 \,\,\,\,\,\, i=1,...,d-1 \ee for the transversal
coordinates of Rindler observers \cite{sanchez-vega}, which is also
valid for inertial coordinates.

In particular, in order to quantize the system and to define the
Hilbert space, we will take a worldsheet foliation in Cauchy
surfaces (topologically $S^1$)
 whose parameter coincides with the proper time of a uniformly accelerated observer in the target space
 time.\footnote{In fact, the worldsheet manifold $W$ may be decomposed as a collection of spatial one
  dimensional manifolds $\Sigma_t \sim S^1 $ and the scalar product is canonically defined by $ (X_1 ,
X_2) = \imath \,\int_{\Sigma_t} \, X_1
\stackrel{\leftrightarrow}{\partial} X_2 = \imath \,\int_{\Sigma_t}
\, X_1
\partial_t X_2 - X_2
\partial_t X_1 $ which allows to quantize the field $X^i $ and to construct the corresponding Hilbert/Fock space.}
In this case we use the Rindler
coordinates $(r,\tau)$ locally induced from the Rindler ones as shown above, and the local string equation we
have to solve is \be
 \left( - r^{-2} \partial_\tau^2 + \partial^2_r + r^{-1} \partial_r \right) X^i =0  \label{eqm}
 \ee
 in each patch (chart) of this type of coordinates in order to cover the worldsheet manifold. Solutions of this are well known \cite{takagi};
then, by considering the proper conditions of smooth matching
between the solutions on each chart, and the periodicity conditions,
we may construct a complete set of eigenfunctions $U_n (\tau, r)$
orthonormal with respect to the scalar product \be ( X_1 , X_2 ) =
\sum_i \, \imath \, \int_i \, X_1
\stackrel{\leftrightarrow}{\partial_{\tau}} X_2 = \sum_i \, \imath
\, \int_{r_i^-}^{r_i^+} dr \, ( \, X_1 \partial_{\tau } X_2 - X_2
\partial_{\tau } X_1\, )\; ,\ee
while the usual string solution in the Minkowski background is
(\ref{stringsol}) (with the Minkowskian scalar product, $ (X_1 ,
X_2) = \imath \,\int_{S^1} d\s \, X_1
\stackrel{\leftrightarrow}{\partial}_t X_2$).

The general solution for closed bosonic string in these coordinates
may be expressed as: \be X^i(\tau , r)= x^i + i \sqrt{\a'/2}
\sum_{\eta}\sum_{n \in Z} \left( b_n^{i \; (\eta)} \, U_n^{*
\;(\eta)}(\tau , r) + {\bar b}_n^{i \; (\eta)} \, U_n^{(\eta)} (\tau
, r)  \right) \label{sr} \ee

In (\ref{sr}) the symbol $\eta= \pmh$ (and $-\eta= \out/\inn $)
takes into account the fact that the worldsheet has a horizon,
  so it is divided into two causally disconnected regions or wedges. Following the standard
  procedure \cite{vitiello}, \cite{takagi}, we can introduce the operators
$d_n^{\eta} =\sum_{n \neq 0}P_n^{\eta}\alpha_n$, ${\bar d}_n^{\eta}
=\sum_{n \neq 0}P_n^{\eta}\b_n$, where $P_n^{\eta}$ is a complete
set of orthogonal functions, and now the operators $b_n^{(\eta)}$
and $d_n^{(\eta)}$ can be related by a Bogoliubov transformation:
\begin{eqnarray}\label{bogol}
b_n^{(\eta)} &=& d_n^{(\eta)} cosh(\epsilon_n) + \bar{d}_n^{(\eta)
\; \dagger}
 sinh(\epsilon_n) =G(\epsilon_n)d^{(\eta)}_n G^{-1}(\epsilon_n), \nonumber \\
\bar{b}_n^{(\eta)\; \dagger}&=& d_n^{(\eta)} sinh(\epsilon_n) +
 \bar{d}_n^{(-\eta) \; \dagger} cosh(\epsilon_n)= G(\epsilon_n)\bar{d}_n^{(-\eta)\;
 \dagger}G^{-1}(\epsilon_n),
\end{eqnarray}
where the coefficients $\epsilon_n$ depend on the coordinate
transformation parameters, then they shall be related to the
acceleration of the observer in the spacetime, and
$G(\epsilon)$ reads: \be G(\epsilon) =  exp\left [
\sum_{\eta}\sum_n \,\theta  \,\,\ep_n\left(
d_n^{\eta}\bar{d}_n^{-\eta}-d_n^{(\eta)\dagger}\bar{d}_n^{(-\eta)\dagger}\right
)\right ] .\ee

It is clear that when the string does not intersect the horizon, the
$G$ transformation  is trivial (it maps $ |0_0\rangle_{out}$ into
itself)\footnote{In general entanglement theory $H= H^\out - H^\inn$
and $[G \, , \, H]=0$; then if the worldsheet metric is horizon
free, the Hamiltonian is $H \equiv \int_{S^1\sim S^\out} h dr =
H^\out $, then we get $[G \, , \,H^\out]=0  $. } . In order to
describe this, we have inserted the parameter $\theta$, defined to
be $1$ if there is a horizon in the
 string worldsheet and $0$ otherwise.

While the operators $d_n^{\eta}$, $\bar{d}_n^\eta$
 annihilate the vacuum state $\left|0_0 \right\rangle= \left|0_0 \right\rangle_{+}
 \left| 0_0 \right\rangle_{-} $, (referred to the two-dimensional
 Minkowski metric $g_0$), the operators
   $b_n^{(\eta)}$, $\bar{b}_n^{(\eta)}$ annihilate the vacuum:

\be  \left| 0(\epsilon) \right\rangle = G(\epsilon)\left|0_0
\right\rangle , \ee that can be written as a $SU(1,1) \times
SU(1,1)$ coherent state:

\be \left | 0(\ep)\right\rangle  = (1/Z)\,\,
 exp\left [\theta \; \sum_{\eta}\sum_n \,(\tanh \ep_n )
 ( d_n^{\dag \, (\eta)} \bar{d}_n^{\dag \, (-\eta)} ) \right] \left|0_0 \right\rangle \,\,\,\,
 ,
 \label{coh}
\ee where \be Z=Z[\ep] \equiv \prod_n cosh^2 \ep_n\, . \ee We
conclude this part by noticing that
the boundary state (\ref{boundary}) may be recovered from the
expression of the fundamental state (\ref{coh}) for a suitable
$\lambda$,  as we should expect. In particular we get: $\tanh
\ep_n = q_n( \lambda)$, and the occupation number of the string
modes is given by $N_n = \sinh^2 \ep_n $, which agrees with a
Bose-Einstein distribution of string modes
 at the temperature $i (4\lambda)^{-1}$. Therefore, the string twist produced by the horizon
 may be interpreted in terms of the temperature/acceleration of the Rindler observer \footnote{
In this case, it shall be thus interpreted as a purely imaginary
twist $\lambda \equiv i \b/4$.}.

\vspace{0.3cm}

Clearly, this {\it vacuum state} consists of a condensate of string modes
placed on the horizon region\footnote{A similar condensed state was
found in \cite{gdn}, when a closed string approaches the null
singularity of the pp-wave time dependent background.}
which describes the critical point where the {\it transition}
 closed (no-entangled) $to$ c-open
(entangled) string happens. Then one may discuss the actual
physical meaning of this transition in a semiclassical language.

In this particular situation, when the closed string (thought here
as a classical extended object) intersects the horizon at some
point, the center of mass of the string is approximately at a
distance $r_c \equiv l_s/2\pi$ from this intersection point, where
the circumference length of the closed string is $l_s$.

On the other hand, the force/acceleration that has to be applied at
the center of mass point in
 order to get a fiducial string\footnote{Since this situation is assumed to be stationary (i.e the string does
not fall into the causally prohibited region), the string has to be
accelerated.} is given by the inverse of its distance (see Ref.
\cite{slips}) to the
 horizon, $a_c \sim 1/r_c$. Then,
 since the order of magnitude of the closed string circumference is
$l_s\sim \sqrt{2\a'}$, we finally conclude  that $a_c = 2\pi / l_s
\sim 2\pi / \sqrt{2\a'} $. Finally this acceleration may be related
to a critical temperature via Unruh effect, since the system at this
point feels a thermal bath
 of temperature $T_c \sim a_c$. So, because the string is an extended object, when the acceleration of
the center of mass point exceeds $a_c$, an event horizon intersects
the worldsheet and degrees of freedom behind it become causally
independent, and the entanglement produced by these hidden degrees
of freedom becomes non trivial. This critical value is indeed
similar to the Hagedorn Temperature, $T_H  \sim 1 / \sqrt{2\a'}$; so
at least, it may be affirmed that \emph{their order of magnitude are
in agreement}. This coincidence suggests a possible identification
of both values and consequently a plausible model for the Hagedorn
transition \footnote{Consisting in the appearing of two (or more) causally disconnected regions in the worldsheet. This should finally depend only on the temperature rather than on a relative magnitude as the acceleration.}, however this is merely an speculation which cannot be
rigorously shown in the simplified context we are considering here.

 Let us conclude this section by pointing out that
  this condensed state exists for accelerated
(fiducial, for black hole horizons) observers so as in the
``membrane paradigm" \footnote{For observers which remain stacionary
with respect to the black hole (and do not fall into), the horizon
region is effectively viewed as a membrane.} \cite{price}, or, in the modern language of
strings and branes, the ``stretched horizon" \cite{suss1,suss2}. So
here, we wish to emphasize
 that this could constitute an appropriate microscopical \emph{model} for this
 object\footnote{For a discussion about the relation of this
  type of coherent states to macroscopical ones, see \cite{ume}.}, which
  may be expressed as a boundary state (eq. (\ref{boundary})) localized in the horizon.

 In fact, it
is believed that the stretched horizon is a hyper-surface, that for
fiducial observers, behaves like an extended object with dissipative
characteristics, a membrane. We are going to show in the next section how a
dissipative behavior appears naturally in the scenario presented
here.

\section{String field and dissipative behavior.}

If the coherent state (\ref {coh}) is assumed to be a microscopic description of the membrane paradigm, which has dissipative attributes \cite{price}, a dissipative behaviour should be then expected to arise in this framework, and as a by-product of the arguments given in the previous section, it should be related to the Hagedorn scale.

The goal of this Section is to argue for the dissipative behavior of
strings in contact with a horizon discussed above when we consider a second quantized description, namely a string field approach. To this purpose
we will only use very basic and generic properties of light-cone
string field theory which are supposed to be valid in this context
\cite{pol1},
 to argue the existence of interaction terms between inaccessible/accessible (in/out) modes in a second quantized Hamiltonian. We will also use some simple arguments of the ``Non Equilibrium Thermofield Dynamics"
\cite{Ari1,Ari2,Ari3}, a unified and canonical formalism which
extends the Thermofield Dynamics (TFD)\cite{ume} to quantum dissipative
systems. According to this, such a term would be an evidence of dissipation, or more weakly: out of equilibrium processes, since an interaction between accessible/inaccessible modes describes energy-momentum exchange among them, which clearly causes lost of microscopic information into the inaccessible system.

In light-cone
string field theory, the interaction diagrams are constructed by demanding  the continuity of worldsheet embedding, which defines a set of gluing equations.  The solution of the gluing equations are boundary states, which are vertex states living in the multi-string Hilbert space. For each vertex state there is a second quantized Hamiltonian which defines the string interaction. This is exactly what we have here.    
We have shown that Rindler observers see the closed string
intersecting the horizon as an entanglement
 of  strings, defined by equation (\ref {coh})(or
(\ref{boundary})).
 On the other hand, this state may be thought as a
squeezed \textit{vertex} state
     $\left|V\right\rangle $ in the 2-string Hilbert space
      $ {\cal H}_\inn \otimes{\cal H}_\out$, corresponding to two c-open strings as discussed
      before. In fact, it is clear that the configuration described above
constitutes a \emph{diagram} at tree level of two strings interacting while
their respective boundaries make contact on the horizon hypersurface and the gluing equations are defined in (\ref{gluing}).
This may be expressed as follows:
 \be 
\left|V\right\rangle= \sum_{\overrightarrow{N}_1,\overrightarrow{N}_2}
C(\overrightarrow{N}_1,\overrightarrow{N}_2) |\overrightarrow{N}_1 ,
\inn \rangle  | \overrightarrow{N}_2 , \out \rangle := | 0(\ep) \rangle \;
,\label{coh-C} \ee
 where the coefficients may be explicitly expressed by \be
C(\overrightarrow{N}_1,\overrightarrow{N}_2) = \langle
 \overrightarrow{N}_1 , \inn | \langle \overrightarrow{N}_2 , \out  | \,\, G \,\, |0_0 \rangle \, , \label{Cexplic}\ee
 where $G$ is given by (\ref{coh}).
 Notice that in each term of equation(\ref{coh-C}) the quantum
numbers $\overrightarrow{N}_1,\overrightarrow{N}_2$ coincide, since
$ C( \overrightarrow{N}_1,\overrightarrow{N}_2 ) = 0 \,\,\,\,
\forall \overrightarrow{N}_1 \neq \overrightarrow{N}_2$. By
projecting these basis states in the position base, we get the wave
functions for the $\overrightarrow{N}$-th level: \be \left \langle x
|\overrightarrow{N}\right\rangle = \prod^{\infty}_{n}
\Psi_{N_n}(x_n) , \ee where we are using the usual notation of
light-cone string field theory, whose oscillation modes shall be
built according to the expansion (\ref{sr}) and the component $N_n$
of the vector $\overrightarrow{N}$ corresponds to the occupation
   number of the oscillator $n$.

   Let us define the  functional field $\Phi^\inn [p_1^+, x_1(\sigma) ,
\sigma \in S^\inn ]$ of the string living in one side of the horizon
and a functional $\Phi^\out [p_2^+ , x_2(\sigma), \sigma \in S^\out
]$ of the string living in the other side. The respective light cone
momenta are $p^+_{1}$ and $p^+_{2}$
and the expansion of the fields in position space is:
 \bea\label{exp}
 \Phi^\inn(p_1^+, x_1(\sigma))&=& \frac{1}{\sqrt{\left|p_1^+\right|}}
 \sum_{\overrightarrow{N}} A^\inn_{\overrightarrow{N}}(p_1^+)\prod^{\infty}_{n}\Psi_{N_n}(x_n),\nonumber \\
 \Phi^\out(p_2^+, x_2(\sigma))&=& \frac{1}{\sqrt{\left|p^+_2\right|}}
 \sum_{\overrightarrow{N}} A^\out_{\overrightarrow{N}}(p^+_2 )
 \prod^{\infty}_{n}\Psi_{N_n}(x_n),\eea
 where the second quantized operators
    $A^\inn_{\overrightarrow{N}_1}(p^+_1)$, $A^\out_{\overrightarrow{N}_2 }(p^+_2)$
     are string creation operators for $p^+ < 0$  and string annihilation operators
      for $p^+ > 0$.

 In a functional representation, the interaction Hamiltonian related to the configuration of the two c-open strings
on the horizon (i.e for Rindler observers) is bilinear in the fields
$\Phi^\inn(p_1^+, x_1(\sigma)$ and  $\Phi^\out(p^+_2, x_2(\sigma))$.
 It may be written in the
number basis as (up to a suitable coupling constant): \be H_I =
\int\, dp^+_1\,d p^+_2
\sum_{\overrightarrow{N}_1,\overrightarrow{N}_2}
C(\overrightarrow{N}_1,\overrightarrow{N}_2)A^\inn_{\overrightarrow{N}_1}(p^+_1)
A^\out_{\overrightarrow{N}_2}(p^+_2) \; ,\label{diss} \ee which
describes the contact interaction between the two strings encoded in
the (vertex) state (\ref{coh-C}), with  the
$C(\overrightarrow{N}_1,\overrightarrow{N}_2)$ coefficients given by
equation (\ref{Cexplic}).

Then, this   Hamiltonian  typically describes a dissipative/out of
equilibrium system; since for observers (which detect the horizon)
in one of these sides, the system/field in the other (hidden) side
may be considered ``non-physical". Here both systems are physically
realized but as observed by Israel \cite{Israel}, the important fact
is that they are causally disconnected. Thus, this structure, with a
coupling between physical and non-physical modes,  is analogous in
form to the Non Equilibrium-TFD, characterized by a Hamiltonian which includes a coupling between
the physical and non-physical modes. It is remarkable that in the
presence of horizons, while quantum fields of particles produces a
scenario similar to equilibrium TFD \cite{Israel,vitiello}, in this
approach, owing to the extended nature of strings, free string
fields behave more according to Non Equilibrium-TFD.

Finally, for completeness, from (\ref{Cexplic}) we may explicitly verify that for
$\theta
 =0$, corresponding to consider the string field in a region distant from the horizon,
  so as for inertial observers:
$C(\overrightarrow{N}_1 ,\overrightarrow{N}_2) \,=\,0 \, , \,
\forall \overrightarrow{N}_1 , \overrightarrow{N}_2 \neq 0$. Thus,
in this case the contact hamiltonian reads \be  \int \,d p^+_1\,  d
p^+_2\, \,\, C(\overrightarrow{0}_1,\overrightarrow{0}_2)\,
A^\inn_{\overrightarrow{0}}(p^+_1) \,
A^\out_{\overrightarrow{0}}(p^+_2) \; ,\label{diss-riv} \ee 
which is irrelevant for
the closed string dynamics observed in inertial coordinates.
 In
fact, this encodes a sort of product between $\Phi^\inn$ and
$\Phi^\out$ that precisely shall be equivalent to a linear term in
the closed string field $\Phi$ which clearly does not contribute to
its own evolution equation.

\section{Concluding remarks}

We proposed a simplified model where the transition, closed $\to$
c-open strings, occurring when the string intersects a horizon (and
the center of mass acceleration is $a_c \sim 2\pi ( l_s)^{-1}$), may
be interpreted in this approach as a limit for the equilibrium of
the system beyond what there is a transition equilibrium/dissipation
rather than a standard phase transition. This critical acceleration
may be related to a temperature via Unruh effect; so the cut
temperature is $T_c \sim 1 / \sqrt{2\a'}$ which remarkably {\it
coincides} with the order of magnitude of the Hagedorn temperature
$T_H$. It seems then to be natural to hypothesize the
possible identification of these critical temperatures and
consequently interpret $T_H$ as being a limit temperature for a
system of \emph{non interacting strings} in equilibrium.

 It is
believed that the Hagedorn regime should be explained in the context
of strong coupling with the background space time since the Hagedorn
singularity arises by summing over highly excited modes in the
partition function. However, this singularity appears \emph{even}
for free strings in a Minkowski space time and in this approach we
precisely get a simplified model where a singularity is also
present, but with some new ingredients related to entanglement and
dissipation, which, rather than to be viewed as an explanation, it
may shed some light on physics in the Hagedorn phase. In fact, we
assumed weakly coupled strings in nearly flat background space times
such that the back-reaction is controlled. This conditions are also
valid for very massive Black Holes, where near horizon gravity is
weak and the metric may be approximated by Rindler
\cite{suss1,suss2}.

Let us remark that this hypothesis would explain why the partition
function corresponding to an {\it equilibrium ensemble} is ill
defined beyond this value (in particular, it diverges). In addition,
these results provide a physical mechanism to enforce the approaches
where the distance of the horizon $( a_c)^{-1}$ constitutes a cut
off in order to evaluate thermal observables (e.g free energy) and
obtain finite results. In fact, closer degrees of freedom could not
be considered in equilibrium and consequently these observables
would be improperly defined in such regions.

Finally, we have also found an horizon filling boundary state, with
many of the properties that one should wish to recover in a
stretched horizon model.

\subsection{Appendix: Induced worldsheet metric in a Rindler
 spacetime}

The string embedding is $X:W\to M$ is assumed to be smooth for
simplicity. $\tau^\mu (\mu = 0..d)$ is the $d$-vector tangent to the
accelerated curve and its affine parameter is $T$, and let $R_\mu
\equiv \partial_\mu R$.

 Let us introduce the vector fields $\tau^a := \nabla^a \tau $, $r^a :=\nabla^a r$
($a= 1, 2$) in $T_p W$, and  define them through the pull-back
 \be \label{emb1}\tau^\mu = \tau^a
\partial_a X^\mu ,\ee
\be \label{emb2}R^\mu = r^a\partial_a X^\mu  .\ee On the other hand,
 the worldsheet metric is consistently induced via this map from the
metric (\ref{rindlertarget})\be
 g_{(W) ab} = \partial_a X^\mu\, \partial_b X^\nu \, g_{(M) \mu \nu}
 ,
 \ee
then the line element reads: \be ds^2=-R^2 d\tau^2 + dr^2 .\ee This
may be verified by virtue of the relations: \be \tau^\mu \tau^\nu
g_{(M) \mu \nu} =\tau^a \tau^b g_{(W) a b} = -R^2 \ee
 \be R^\mu R^\nu g_{(M)
\mu \nu} =r^a r^b g_{(W) a b} = 1 \ee
 \be \tau^\mu R^\nu g_{(M)
\mu \nu} = \tau^a r^b g_{(W) a b} = 0 .\ee Finally, by using
(\ref{emb2}) and the definition of the Rindler coordinates (eq.
(\ref{rindlertarget})), we can notice that $\partial_\tau  R =\tau^a
\partial_a R = \tau^a (\partial_a X^\mu \,\partial_\mu R) =\tau^\mu
\partial_\mu R = \frac{\partial R}{\partial T}=0$. This implies that $R=R(r)$ and it may be chosen
as $R(0)=0$, thus near of $r=0$ the worldsheet metric takes the form
(\ref{rindler}): \be\label{rindlerk} ds^2 = - \kappa r^2 d\tau^2 +
dr^2 \,\,\,  \kappa =constant .\ee
and $\kappa$ is a positive constant, due to the signature, which may be taken to be one.

\section{Acknowledgements}
M. C. B. Abdalla and M. Botta C. acknowledge CNPq and CONICET respectively for financial support. Special thanks are due to Alexandre Gadelha and D\'afni Marchioro for
fruitful discussions.
The authors would like to thank also to IFT-UNESP (Sao Paulo, Brasil),
where this work was initiated.

\end{document}